\documentclass[12pt]{article}
\usepackage{subfigure}
\usepackage{graphicx}
\usepackage{amssymb,amsmath,mathrsfs,bm}
\numberwithin{equation}{section}
\newcommand{\newc}{\newcommand}
\newc{\ra}{\rightarrow}
\newc{\lra}{\leftrightarrow}
\newc{\be}{\begin{equation}}
\newc{\ee}{\end{equation}}
\newc{\bg}{\begin{gathered}}
\newc{\eg}{\end{gathered}}
\newc{\bs}{\begin{split}}
\newc{\es}{\end{split}}
\newc{\ba}{\begin{eqnarray}}
\newc{\ea}{\end{eqnarray}}
\newc{\ov}{\overline}
\newc{\pa}{\partial}
\newc{\D}{\Delta}
\begin{document}

\begin{titlepage}
\begin{center}
\begin{Large}
{\bf
The quantum cat map on the modular discretization of  extremal black hole horizons

}

\end{Large}

\vskip1.2truecm

{\large{Minos Axenides$^{1}$, Emmanuel Floratos$^{1,2}$ and Stam Nicolis$^{3}$}}

{\sl 
$^{1}$ NCSR ``Demokritos'', Institute of Nuclear and Particle Physics\\
15310 Aghia Paraskevi, Attiki, Greece \\
$^{2}$ Physics Department, University of Athens, Zografou University Campus\\
15771 Athens, Greece\\
$^{3}$ CNRS--Institut Denis Poisson (UMR 7013)\\
Universit\'e  de Tours, Universit\'e d'Orl\'eans\\
Parc Grandmont, 37200 Tours, France
}
\vskip1truecm

{\tt E-Mail: axenides@inp.demokritos.gr, mflorato@phys.uoa.gr, stam.nicolis@idpoisson.fr}
\end{center}
\begin{abstract}

\vskip0.8truecm

Based on our recent work on  the discretization of the  radial  AdS$_2$   geometry of extremal BH horizons,we present a toy model for the chaotic unitary evolution of infalling single particle wave packets.

We construct explicitly the eigenstates and eigenvalues for the single particle dynamics for an observer  falling into the BH horizon, with time evolution operator the quantum Arnol'd cat map (QACM).

Using these results we  investigate the validity of  the eigenstate thermalization hypothesis (ETH), as well as that of the fast scrambling time bound (STB).

We find that the QACM, while possessing a linear spectrum, has  eigenstates, which  are  random and satisfy the assumptions of the ETH.

We also find that the  thermalization of infalling wave packets  in this particular model  is exponentially fast, thereby   saturating  the STB, under the constraint that the finite dimension of the single--particle Hilbert space takes values in the set of Fibonacci integers.
\end{abstract}
\end{titlepage}
\section{Introduction}\label{intro0}

A very interesting revival of the old relation between the near horizon shock wave BH geometries with gravitational memory effects and the information paradox has recently appeared~\cite{hooft,strom}.

It seems  possible in principle, that the horizon region of BH could form a random  basis of purely geometrical data  of all of its past and recent  history, through the 't Hooft mechanism of permanent space-time displacements caused by high energy scattering events of infalling  wave packets~\cite{hooft}.

In the language of refs.~\cite{strom},  such data can be identified with  the soft  hair of the BH, whose origin is the infinite number of conservation laws, described by the BMS group.

This new reincarnation of the 't Hooft-Susskind horizon holography provides a new framework to study mechanisms by which   past and recent  memories of  the shock-wave spacetime geometry structure are  encoded in the angular and time correlations  of  the emitted Hawing radiation.

Quite recently the non--unitarity of the Hawking radiation has been interpreted as the result of integrating all the BMS soft graviton cloud, accompanying the hard Hawking quanta~\cite{hardsoft}. 

Although a realistic calculation with a truly chaotic horizon region, in general,  still isn't possible,  some progress could be made using simple mathematical  toy models, which can describe how the information carried by  infalling wave packets, is scrambled,  through a chaotic single--particle S-matrix~\cite{barrabes_etal,papadodimas,stanford_susskind,shenker_stanford,mezei_stanford,polchinski}.

Proposals for a chaotic, discretized, dynamics for the microscopic degrees of freedom of the stretched horizon have been discussed for quite some time in the literature~\cite{sussk,BFKS,polchinski08,avery,magan}.

So it is imperative to separate the issue of the chaotic dynamics and  geometry of the near horizon region from any issues regarding Hawking radiation. 

It is possible to do so, when studying probes of extremal black hole horizons; for, in that case, since the Hawking temperature vanishes, while the entropy does not, there is no Hawking radiation. 

In this case, it is known that the near horizon geometry is described by a metric, that factorizes into a radial--temporal part that can be identified as an AdS$_2$ manifold, while the angular part describes the charges. 

It is possible to construct a model for random AdS$_2$ geometries, inspired by the shock wave geometries, necessary for the chaotic dynamics of the probes, by introducing  a consistent discretization of the AdS$_2$ near horizon geometry of nearly extremal black holes. This is the so--called the modular discretization, AdS$_2[p]$, for every prime  integer $p$~\cite{AFNAdS2CFT1}. In this framework, the entropy of the black hole, is identified with the Kolmogorov--Sinai entropy of the deterministic, chaotic,  dynamics of the geometric, microscopic,  degrees of freedom, defining the near horizon geometry. 

AdS$_2[p]$ is a specific discrete  deformation of its continuous counterpart.It  has a random structure due to the modular arithmetic. As explained in ref.~\cite{AFNAdS2CFT1}, by  explicit calculation, this specific discretization is chosen among many possible discretizations because it provides a way of constructing an holographic correspondence  between the bulk, AdS$_2[p]$ and its
boundary $\mathbb{RP}^1[p]$, the discrete  projective line.

 The reason  this discrete holography exists at all is that it is possible to realize the action of  the discrete and  finite symmetry group of AdS$_2[p]$, which is PSL$_2[p]$:  it  acts as an isometry group of the bulk and as the (M\"obius) conformal  group on the boundary.

 This discrete geometry provides also a natural  framework for describing the  single particle dynamics, via observers, with time evolution operators that are elements of the isometry group. This is a discrete analog  of the superconformal quantum mechanics  of probes near the horizons of large extremal black holes~\cite{townsendetal}.

In the present work we  specify the infalling, accelerating, observer by   the well known Arnol'd cat map (ACM)~\cite{arnoldcatmapbooks}. 
This map defines a, particular, observer, ACM,  who, by performing single--particle scattering on the horizon of the black hole, can probe the randomness of the geometry and it is consistent with the isometries of the background  
since it belongs to the discrete isometry group of AdS$_2[p]$. The discreteness of the geometry implies that the global coordinates of AdS$_2[p]$ are discrete.
On the other hand, the time, measured by the ACM observer, is the iteration step of the corresponding map. 

In this work we study the quantum dynamics of the probe in this discrete, background geometry. 

It is important to stress that {\em both}, probe and background geometry, have a finite dimensional space of states. What we study is how superpositions of the states of the probe evolve, when the background geometry is found in any given, fixed state. 

The ACM can be ``quantized'', i.e. it is possible to define a $p\times p$ unitary evolution operator, called the quantum Arnol'd cat map (QACM). This definition uses  the Weil representations  of SL$_2[p]$ and especially those that correspond to its  projective action, by  PSL$_2[p]$ on AdS$_2[p]$. This construction extends the results for the case of the  discrete torus~\cite{quantum_maps,knabe,floratos89,athanasiu_floratos,fastqmaps}.

An introduction to the requisite tools from arithmetic geometry and computational number theory can be found in ref.~\cite{arithmetic_geometry}.

Next we proceed with  the plan of the paper:

In section~\ref{ArnFibChaos} we recall the properties of  the ACM, its relation to the Fibonacci sequence and its periods mod$\,p$. We study its group of symmetries, inside PSL$_2[p]$, i.e. the set of elements of PSL$_2[p]$, that commute with it.

In section.~\ref{QACM} we construct explicitly the exact Quantum Arnol'd cat map, using the metaplectic (or Weil) representation of SL$_2[p]$ which is reducible and splits into $(p+1)/2$ and $(p-1)/2$ dimensional irreducible ones. One of these two unitary irreps, depending on the form of the prime number $p$, is also a representation of PSL$_2[p]$. 
This particular representation defines, for every prime, $p$, the Hilbert space of states of the infalling wavepackets on AdS$_2[p]$, while the other one is appropriate for the case of the torus.

 We determine analytically the spectrum and the eigenstates of the QACM and we compute  their degeneracies. 

We find  the interesting result that, while  the spectrum is
linear,  the eigenstates are chaotic in a very specific way, that is, the squares of the absolute  values of the amplitudes (probabilities) are drawn from a (discrete) Gaussian distribution, while their phases have a flat distribution. 
These results are known to be the premises  for  the eigenstate thermalization hypothesis (ETH) for quantum ergodicity  or unitary thermalization~\cite{ETH}.

In section~\ref{FibhorizonsSTB} we review the Eigenstate Thermalization Hypothesis and we stress that its premises can be checked to hold within ou our framework. 
We use the results obtained in section~\ref{QACM} to study the spectrum of scrambling times  and we find that for Fibonacci integer values the scrambling time bound of Susskind and Sekino is saturated.

Finally in section~\ref{Concl} we discuss our results and open problems for future work.

 In appendix  A  we collect all the necessary material for the detailed construction of the Weil representation of  SL$_2(p)$ and PSL$_2(p)$ and 
 we present the technical details for the analytic construction of the eigenstates and eigenvalues of the QACM.

%===============================================================================================
\section{ The Arnol'd cat map and Fibonacci chaos on AdS$_2[N]$}\label{ArnFibChaos}
%===============================================================================================

We review the description of the modular discretization of AdS$_2$~\cite{AFNAdS2CFT1}.

We define the modular discretization  by  replacing  the set of real numbers, $\mathbb{R}$, by the set of integers modulo $N$. The so obtained  coset finite geometry  AdS$_2[N] =SL_2(\mathbb{Z}_N)/SO(1,1,\mathbb{Z}_N)$ 
is a discrete deformation of its continuous counterpart, AdS$_2[\mathbb{R}]=SL_2(\mathbb{R})/SO(1,1,\mathbb{R})$.
AdS$_2[N]$ is a finite and  random set of points in the embedding Minkowski spacetime. When $N$ is prime this is an ``arithmetic geometry'' in the mathematical literature~\cite{arithmetic_geometry}, that is,  a geometry  over a finite field.

This discretization can be used to describe  nonlocality, chaos and quantum information processing in the vicinity of the BH horizon, as well as defining a discrete version of the AdS$_2$/CFT$_1$ holography.

\begin{figure}[thp]
\begin{center}
\subfigure{\includegraphics[scale=0.5]{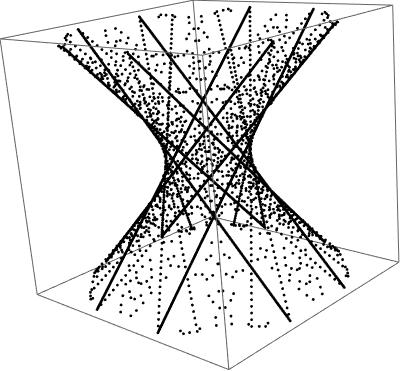}}
\subfigure{\includegraphics[scale=0.5]{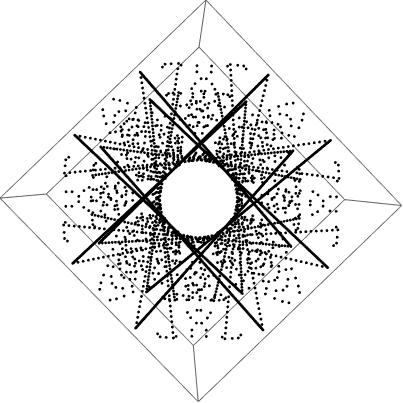}}
\end{center}
\caption[]{The rational points on AdS$_2[p]$--side view and top view, for $N=47, M=48, L=144$.}
\label{AdS2rp}
\end{figure}

The set of points of the finite geometry of AdS$_2[N]$ is, by definition, the set of integer solutions $\mathrm{mod}\,N$ of the equation
\begin{equation}
\label{AdS2_p}
x_0^2 + x_1^2 -x_2^2\equiv\,1\,\mathrm{mod}\,N
\end{equation}
This set--an example of which is shown in fig.~\ref{AdS2rp}--is constructed by noting that, for any triplet of integers, $(k,l,m)$, that satisfy eq.~(\ref{AdS2_p}) mod $N$, there exists an integer $M\equiv\,1\,\mathrm{mod}\,N$, such that 
the triple of rational points $(k/M,l/M,m/M)\equiv(x_0,x_1,x_2)$,  satisfies the equation $x_0^2+x_1^2-x_2^2=1$, i.e. it defines rational points of the continuum
AdS$_2$ manifold. 

If we fix an ``infrared cutoff'', $L$, for $x_2$, $|x_2|\leq L$, by increasing the denominator $M$, we can obtain in this way a rational approximant to the  continuum AdS$_2$ geometry.

A straightforward prescription for constructing all the solutions of eq.~(\ref{AdS2_p}), i.e. the points of AdS$_2[N]$, 
is as follows:

\begin{equation}
\label{LCAdS2N}
\begin{array}{l}
\displaystyle
x_0\equiv (a-b\,\mu)\,\mathrm{mod}\,N\\
\displaystyle
x_1\equiv  (b+a\,\mu)\,\mathrm{mod}\,N\\
\displaystyle
x_2\equiv\,\mu\,\mathrm{mod}\,N
\end{array}
\end{equation}
where $a^2 + b^2\equiv\,1\,\mathrm{mod}\,N$ and $a,b,\mu\in\{0,1,2,\ldots,N-1\}$~\cite{AFNAdS2CFT1}.

Thus, the discretized, spatial part, along $x_2$, consists of $N$ points and the Hilbert space of single--particle states has dimension $N$.  The global, AdS$_2$, time, is parametrized by the points of the discrete circle, $a^2 + b^2\equiv\,1\,\mathrm{mod}\,N$. The proper time of the ACM observer is identified with the number of iterations  of the  ACM mod $N$. 
Due to the mod $N$ arithmetic, the  global, AdS$_2$ time and the proper time of the ACM are periodic. 

If we assign, to each spatial point, a two--state system, the microscopic degrees of freedom of the near--horizon discrete geometry, have a Hilbert space of dimension$\,\propto\,2^N$. From this we conclude that the entropy, $S_\mathrm{BH}$,  of such configurations, is proportional to $\log 2^N=N\log 2$, which is the number of spatial points.

The discussion of the group theoretic properties of this discrete geometry is facilitated, if we restrict $N$ to be a prime integer, $p$. The extension for arbitrary, odd, integer values of $N$ is easily realized by using appropriate factorization theorems~\cite{fastqmaps}. 

The finite geometry, AdS$_2[p]$, has as isometry group the finite projective modular group, PSL$_2[p]$. This group is obtained as the reduction mod$\,p$, of all elements of $PSL(2,\mathbb{Z})$. The kernel of this homomorphism is the ``principal congruent subgroup'', $\Gamma_p$. The order of PSL$_2[p]$ is $p(p^2-1)/2$ and the order of its dilatation subgroup is $(p-1)/2$, thus, the number of points of AdS$_2[p]$ is $p(p+1)$.

It is easy to find the number of points of AdS$_2[N]$, for any integer $N$.

 Numerical experiments suggest the following recursion relation for the number of points of AdS$_2[p^k]$, ${\sf Sol}(p^k)$,
\begin{equation}
\label{solpk}
{\sf Sol}(p^k)=p^{2(k-1)}{\sf Sol}(p)\Rightarrow {\sf Sol}(p^k)=p^{2k-1}(p+1)
\end{equation}
where ${\sf Sol}(p)=p(p+1)$ and $k=1,2,\ldots$ for any prime integer $p$.

For $N=2^n$ we find ${\sf Sol}(2)=4$, ${\sf Sol}(4)=24$, and ${\sf Sol}(2^k)=4{\sf Sol}(2^{k-1})$, for $k\geq 3$. We remark that $N=4$ is an exception. 
The solution is ${\sf Sol}(2^k)=2^{2k+1}$, for $k\geq 3$. 

Therefore we may deduce  the expression for the number of points, for any integer $N$ by prime factorization. 

From these results we deduce that, for large $N$, the number of solutions, mod$\,N$, scales like the area, i.e. $N^2$. So most of the points of AdS$_2[N]$ are close to its boundary and holography is possible in this case too~\cite{witten_susskind}.

Next we discuss the discrete time evolution of the motion of a particle on AdS$_2[N]$. To every point $x_\mu\in\mathrm{AdS}_2[N]$, where $\mu=0,1,2$,  can be assigned a traceless, $2\times 2$, matrix ${\sf X}$ 
\begin{equation}
\label{Weyl}
{\sf X}\equiv\left(\begin{array}{cc} x_0 & x_1+x_2 \\x_1-x_2 & -x_0\end{array}\right)
\end{equation}
whose determinant is,
 $\mathrm{det}\,{\sf X}=-x_0^2-x_1^2+x_2^2\equiv-1\,\mathrm{mod}\,N$.

The discrete time evolution, for an observer, defined by its evolution matrix,  ${\sf A}\in \mathrm{PSL}_2[N]$, is given by the recursion relation
\begin{equation}
\label{mapiterA}
{\sf X}_{n+1}={\sf A}{\sf X}_n{\sf A}^{-1} = {\sf A}^n{\sf X}_0{\sf A}^{-n}\,\mathrm{mod}\,N
\end{equation}
where  $n=0,1,2,\ldots$ labels the (stroboscopic) time of the  observer and ${\sf X}_0$ is the initial point of the trajectory.

We make a specific choice, introduced in ref.~\cite{AFNAdS2CFT1}, of an observer,  described by the Arnol'd cat map(ACM):
\begin{equation}
\label{Acatmap}
{\sf A}=\left(\begin{array}{cc} 1 & 1\\ 1 & 2\end{array}\right)
\end{equation}
The map corresponds to successive kicks, forwards and backwards along the light cone of AdS$_2[N]$, since 
\begin{equation}
\label{cat_translations}
{\sf A} = \left(\begin{array}{cc} 1 & 0 \\ 1 & 1\end{array}\right)\left(\begin{array}{cc} 1 & 1 \\ 0 & 1\end{array}\right)={\sf L}{\sf R}^{-1}
\end{equation}
and is an element of $\mathrm{PSL}_2(\mathbb{Z}_N)$.  We point out that this action differs from the action of ACM  on the torus, which is linear in ${\sf A}$~\cite{arnoldcatmapbooks}.

We choose to use this  particular map, for the following reasons:

 \begin{itemize}

\item 
The ACM has been thoroughly studied for its area preserving  action on the classical toroidal phase space and it is  known to possess  ergodicity, exponentially fast mixing and an infinite number of unstable periodic orbits.The important property of mixing, which is the technical definition of scrambling, assumes that the phase space is compact.

The ACM, acting on AdS$_2[N]$, according to~(\ref{mapiterA}), induces a discrete Lorentz transformation. The mod $N$ prescription guarantees mixing and, thus, ergodicity of the dynamics. 

%\item 
%The interesting feature of the toroidal periodic orbits is that they are all known. Indeed all the rational points of the torus, with common denominator $N,$ are mapped onto themselves under the action of the ACM, which then has a definite period, $T(N),$ depending randomly on $N$~\cite{falk_dyson}.
%
%These are the only unstable periodic orbits of ACM and, as $N$ goes to infinity, we cover all the rational points of the torus and finally we reach the continuum and the, truly, chaotic orbits.

\item 
The mod $N$ prescription provides chaotic orbits on AdS$_2[N]$ but, since ACM has a finite period, $T(N)$, depending, randomly, on $N$, all of these orbits are, also, periodic. Their chaotic nature can be seen up to evolution time $\leq T(N)/2$. 
\item 
The ACM has been studied intensively also as a toy model for semiclassical quantum chaos on the toroidal phase space~\cite{quantum_maps}, although the degeneracies in its spectrum impose  additional constraints on its quantum ergodic properties~\cite{quantum_ergodicity}.

Here we extend the study to  the classical and quantum motion of particles under the Arnol'd cat map on the discretized AdS$_2[N]$ geometry.

As we pointed out above, this is a discretized deformation of the continuous,  radial and time, geometry  of the near horizon region of extremal BHs~\cite{bh_entropy}. 

The motion we study is the longitudinal motion of probes and it differs  from the motion along the horizon of the black hole, which is the two dimensional sphere. On the other hand, physically, the scrambling of information on the horizon happens at the same time as the longitudinal (radial) scrambling~\cite{silverstein}.
\end{itemize}

 An important property of  ACM is that it is  known to generate  the sequence of Fibonacci numbers, $f_n$, with $n=1,2,\ldots$, defined by 
\begin{equation}
\label{fibonaccimatrix}
\left(\begin{array}{c} f_n\\f_{n+1}\end{array}\right) = 
\left(\begin{array}{cc} 0 & 1 \\ 1 & 1\end{array}\right)
 \left(\begin{array}{c} f_{n-1}\\f_n\end{array}\right)
\end{equation}
where $f_0=0$ and $f_1=1$. 

We observe that 
\begin{equation}
\label{falk_dyson_decomp}
{\sf A} = \left(\begin{array}{cc} 1 & 1 \\ 1 & 2\end{array}\right)=
\left(\begin{array}{cc} 0 & 1\\ 1 & 1\end{array}\right)^2
\end{equation}
therefore 
\begin{equation}
\label{falk_dyson2}
{\sf A}^n = \left(\begin{array}{cc} f_{2n-1} & f_{2n}\\ f_{2n} & f_{2n+1}\end{array}\right)
\end{equation}
with $n=1,2,3,\ldots$

Falk and Dyson~\cite{falk_dyson} studied the periods, $T(N)$, of the iteration
\begin{equation}
\label{modN}
{\sf A}^n\,\mathrm{mod}\,N=\left(\begin{array}{cc} f_{2n-1} & f_{2n}\\ f_{2n} & f_{2n+1}\end{array}\right)\,\mathrm{mod}\,N
\end{equation}
for various classes of the integers $N$. 

\begin{figure}[thp]
\begin{center}
\includegraphics[scale=0.8]{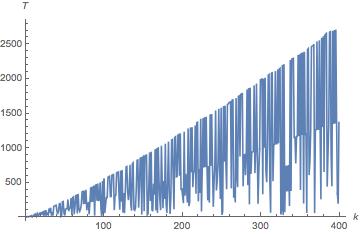}
\end{center}
\caption[]{The period, $T$,  of the Arnol'd cat map as a function of  the order, $k$, of the prime $p_k$, for the first 400 primes.}
\label{fibperiod}
\end{figure}

$T(N)$ is the smallest, positive, integer, such that 
\begin{equation}
\label{periodfibN}
\begin{array}{l}
\displaystyle
{\sf A}^{T(N)}=\left(\begin{array}{cc} 1 & 0\\ 0 & 1\end{array}\right)\,\mathrm{mod}\,N\\
\end{array}
\end{equation}
Thus $T(N)$  is, also, the period of the Fibonacci sequence mod $N$, which  is known to be a ``random'' function of $N$--cf. fig.~\ref{fibperiod}.

We now turn to the discussion of the scrambling time, $t_\mathrm{scrambling}$,  of the black hole horizon geometry, which has been introduced in~\cite{scrambling} as the time necessary for uniform spreading of the  distribution of the microscopic degrees of freedom, near  the black hole horizon, under an external perturbation. Here we use the probe approximation for an infalling, single--particle,  wavepacket as such a perturbation  and we assume that the scrambling time of the black hole horizon is the same as that of the wavepacket. For a Gaussian wavepacket the scrambling time is the time necessary for its uniform spreading along the horizon. Moreover, we focus on the radial dependence of the spreading, for which the scrambling time is the same as that of the transverse spreading. 

More technically, the scrambling time, $t_\mathrm{scrambling}$, as defined above, is identical with the mixing time, $t_\mathrm{mixing}$ of the  dynamical system, here the cat map,  on AdS$_2[N]$~\cite{arnoldcatmapbooks}. 
Since $T(N)$ is the period of ${\sf A}\,\mathrm{mod}\,N$, the maximum available time for scrambling  is proportional to $T(N)/2$, therefore
\begin{equation}
\label{mixingtime}
t_\mathrm{scrambling}=t_\mathrm{mixing}\leq\frac{T(N)}{2}
\end{equation}
for all $N$. 

From the work of Falk and Dyson~\cite{falk_dyson},  if $N=f_{2k}$, then $T(N)=2k$. Therefore, $t_\mathrm{scrambling}=t_\mathrm{mixing}\leq k$. We recall that the solution of the Fibonacci recurrence is given by
\begin{equation}
\label{fibonrecursion}
f_n = \frac{1}{\sqrt{5}}\left( \lambda^n -\frac{(-)^n}{\lambda^n}\right)
\end{equation}
with 
\begin{equation}
\label{eigenvaluesACT}
\lambda = \frac{1+\sqrt{5}}{2}
\end{equation}
known as the Golden Ratio. For $n=2k>>1$, $f_{2k}=N \sim (1/\sqrt{5})\exp(2k\log\lambda)$, so $T(N)\sim\ln N$. These orbits mod $f_{2k}$ are ``short'' orbits and, in order to get mixing, we have to take ``large'' values of  $k$. At the quantum level, the role of ``short'' orbits has been connected with that of ``scars''~\cite{quantum_ergodicity}.

 Such  ``short'' periods of the ACM imply  the existence of non--trivial conservation laws, that is, elements of $SL(2,\mathbb{Z}_N)$, that commute with it. These form an  abelian  group, the commutant, $G({\sf A})$. For prime values of $N$, it is cyclic, i.e. there exists a ``primitive element'', whose powers generate all the others.  Among the elements of this group, obviously, are the powers of ${\sf A}$ mod$\,N$; the non--trivial conservation laws are described by the complement thereof. The general element of $G({\sf A})$ has the form
\begin{equation}
\label{commutantACM}
{\sf C}(k,l)=\left(\begin{array}{cc} k & l\\ l & k+l\end{array}\right)
\end{equation} 
with $k,l$ integers, satisfying  the constraint $k^2 +kl-l^2=1$. This can be cast in the form of Pell's equation 
\begin{equation}
\label{Pell}
x^2-5y^2=1
\end{equation}
with $x=k+(l/2)$ and $y=l/2$, in which case  $l$ must be even. The ``trivial'' conservation laws are given by the Fibonacci numbers, $k=f_{2n-1}$, $l=f_{2n}$, for all $n$; in this case, integer solutions of Pell's equation correspond to $n=3m$, with $m=1,2,\ldots$~\cite{arithmetic_geometry}. 

For prime values of $N$ the period of the ACM divides the period of the commutant. If the two periods are equal, the ACM is a primitive element of  $G({\sf A})$ and there aren't any non-trivial conservation laws. If they're not, then the ACM is a power of the primitive element of the commutant. This power determines the degeneracies of the {\em quantum} ACM, as we shall see in the next section. 

The deterministic, chaotic, orbits of ACM on AdS$_2[N]$ can be obtained as follows: 
If we take as initial point  ${\sf X}_0\equiv(x_0,x_1,x_2)$, we  find the corresponding sequence, $\{{\sf X}_n\}$,
\begin{equation}
\label{mapiterA_solX}
\begin{array}{l}
\displaystyle
{\sf X}_n \equiv \left(\begin{array}{cc}  x_0^{(n)}   &  x_1^{(n)}+x_2^{(n)} \\ x_1^{(n)}-x_2^{(n)}   & -x_0^{(n)} \end{array}\right)=
{\sf A}^n {\sf X}_0 \left[{\sf A}^{-1}\right]^n =\\
\displaystyle
\left(\begin{array}{cc} f_{2n-1} & f_{2n}\\ f_{2n} & f_{2n+1}\end{array}\right)
\left(\begin{array}{cc}  x_0   &  x_1+x_2 \\ x_1-x_2   & -x_0 \end{array}\right)
\left(\begin{array}{cc} f_{2n+1} & -f_{2n}\\ -f_{2n} & f_{2n-1}\end{array}\right)\,\mathrm{mod}\,N
\end{array}
\end{equation}
Writing out the results we find that $(x_0^{(n)},x_1^{(n)},x_2^{(n)})$ is given by the action of
 elements ${\sf L}_n\in SO(2,1)$, with integer coefficients mod $N$, that act on the initial point $(x_0,x_1,x_2)$, for every time step $n$:
\begin{equation}
\label{Lorentz_n}
 \begin{small}
 {\sf L}_n\equiv
 \left(\begin{array}{ccc}  
 \left(f_{2 n}\right){}^2+f_{2 n-1} f_{2 n+1} & f_{2 n} f_{2 n+1}-f_{2 n} f_{2 n-1} & -f_{2 n} f_{2 n-1}-f_{2 n} f_{2 n+1} \\
 f_{2 n} f_{2 n+1}-f_{2 n} f_{2 n-1} & -\left(f_{2 n}\right){}^2+\frac{1}{2} \left(f_{2 n-1}\right){}^2+\frac{1}{2} \left(f_{2 n+1}\right){}^2 & \frac{1}{2} \left(f_{2
   n-1}\right){}^2-\frac{1}{2} \left(f_{2 n+1}\right){}^2 \\
 -f_{2 n} f_{2 n-1}-f_{2 n} f_{2 n+1} & \frac{1}{2} \left(f_{2 n-1}\right){}^2-\frac{1}{2} \left(f_{2 n+1}\right){}^2 & \left(f_{2 n}\right){}^2+\frac{1}{2} \left(f_{2
   n-1}\right){}^2+\frac{1}{2} \left(f_{2 n+1}\right){}^2 \\
 \end{array}\right)
 \end{small}
 \end{equation}

It is noteworthy that ${\sf L}_n$ has fractional coefficients, which means, even without applying the mod operation, that the corresponding combinations of integer coordinates on the hyperboloid are even. Also, that, for even $N$, these matrices must be defined separately.

The relevance of these expressions is that they highlight the, classical, deterministic, chaotic dynamics of the ACM on AdS$_2[N]$.

\begin{figure}[thp]
\includegraphics[scale=0.6]{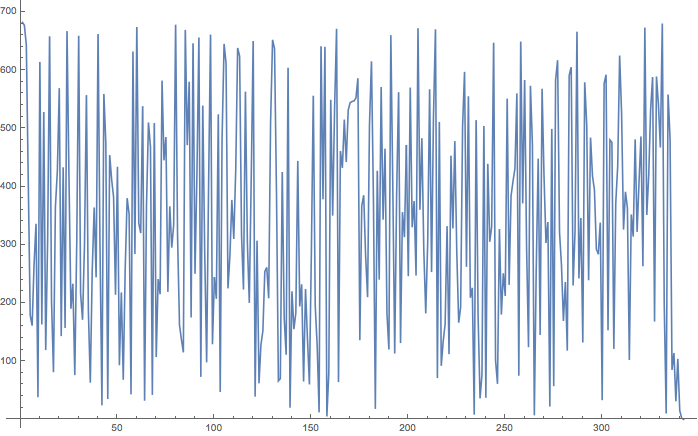}
\caption[]{$x_2^{(n)}$ mod 683 as a function of the time step $n$. The period is found to be equal to $684=N+1$.}
\label{X2nmod}
\end{figure}

Indeed, the motion of a particle under ACM, along the spatial direction, $x_2$, of AdS$_2[N]$ can be seen to be fully chaotic and mixing, cf. fig.~\ref{X2nmod}.

%=====================================================================================================
\section{Chaotic eigenstates of the QACM on AdS$_2$}\label{QACM}
%====================================================================================================
Quantum mechanics for a local  observer in AdS$_2$ is defined, once a choice of time evolution has been made. As discussed in the previous section the  isometry groupμ  $SL(2,\mathbb{R})$, can be used to identify the time evolution operator for an observer with an element of this group. 

Since we have discretized the geometry locally, the canonical variables, for  any observer, are the exponentials of position and momentum operators, that define the generators of the finite Heisenberg--Weyl group, HW$_N$. 

The classical isometry of the discretized geometry is $PSL(2,\mathbb{Z}_N)$. Having chosen as time  evolution map, ${\sf A}$,  the Arnol'd cat map, we have, in fact,  specified the observer and its time is defined as the number of iterations of this map.  We can now construct the corresponding quantum evolution map, QACM, choosing, for simplicity, $N=p$ prime. This will be, also, the dimension of the single--particle Hilbert space of this observer.

 The unique Hilbert space, for all observers, is defined by the irrep of the Heisenberg--Weyl group, as we discussed in ref.~\cite{AFNAdS2CFT1}. Every observer in the bulk can reconstruct the algebra of his/her observables from those of the conformal field theory (for AdS$_2$, this is conformal quantum mechanics) on the boundary using bulk--to--boundary Green functions. 
 
 To construct the unitary (quantum) evolution operator, $U({\sf A})$, corresponding to the, classical, Arnol'd cat map, ${\sf A}$, we shall use  the Weil representation of  PSL$_2[p]$.

The detailed construction of $U({\sf A})$ is given, for completeness, in appendix~\ref{UofA}, for the group SL$_2[p]$.  This representation, by construction, is the direct sum of two, irreducible, representations, of dimensions $(p+1)/2$ and $(p-1)/2$. Since the action of ${\sf A}$ on AdS$_2$ does not distinguish the action of ${\sf A}$ from that of $-{\sf A}$, it realizes a projective action--which is how 
 quantum mechanics on AdS$_2$ differs from the torus, $\mathbb{T}^2$. Therefore it is necessary to choose one of these two representations, thereby imposing the constraint that it is, also, a representation of PSL$_2[p]$. 
 
 One important property of the quantization procedure is that, for any two elements, ${\sf A}_1,{\sf A}_2\in\mathrm{PSL}_2[p]$, $U({\sf A}_1{\sf A}_2)=U({\sf A}_1)U({\sf A}_2)$. This implies the interesting fact that, to calculate the quantum evolution, at time $n=1,2,3,\ldots$, it suffices to compute $U({\sf A}^n)$, which is equal to $\left[U({\sf A})\right]^n$--realizing  a very big simplification in the calculation of time correlation functions.  Therefore, the period of the quantum map is equal to that of the classical map and this determines the degeneracies of the spectra and the conservation laws. 
 
 In the following we sketch the main steps of the construction of the  eigenstates and eigenvalues of the QACM. 
 
We remark  that the restriction to prime values of $N$  for which  5 is a quadratic residue mod $N$ will make possible the analytic construction of eigenstates and eigenvalues of the QACM. It appears that,  up to now, explicit expressions for  the  eigenstates and eigenvalues of the QACM  are not known for generic $N$~\cite{quantum_maps}; so our results, for these primes, are new. 

 The basic idea comes from the observation  that the classical ACM can be diagonalized over the finite field $\mathbb{F}_p=\{0,1,2,\ldots,p-1\}$, if 5 is a quadratic residue mod $p$.   So to avoid unnecessary technical  complications, we choose the prime $p$ to be of the form $4k-1$, since, in that case, if  5 is a quadratic residue $\mathrm{mod}\,p$ it is easy to  construct $\sqrt{5}\,\mathrm{mod}\,p$. (If 5 isn't a quadratic residue mod $p$, we must work in the corresponding quadratic extension.)

We can check, in this case,  that $a\equiv 5^k\,\mathrm{mod}\,p$ satisfies $a^2\equiv 5\,\mathrm{mod}\,p$.  The eigenvalues of ${\sf A}$ are then,
\begin{equation}
\label{eigenvaluesA}
\lambda_\pm\equiv\frac{3\pm a}{2}\,\mathrm{mod}\,p
\end{equation} 
Moreover, there is an element, ${\sf R}\in\mathrm{SL}_2[p]$, that diagonalizes ${\sf A}$, 
\begin{equation}
\label{eigenvectorsA}
{\sf A}= {\sf R}{\sf D}_{\sf A}{\sf R}^{-1}
\end{equation}

where ${\sf D}_{\sf A}=\mathrm{diag}\left(\lambda_+,\lambda_-\right)$. We can deduce that

\begin{equation}
\label{UAR}
U({\sf A}) = U({\sf R})U({\sf D}_{\sf A})U({\sf R})^\dagger
\end{equation}
and that $U({\sf D}_{\sf A})$ is the circulant matrix 
\begin{equation}
\label{circulant}
\left\langle l|U({\sf D}_{\sf A})|k\right\rangle = \delta_{\lambda_+ k,l}=\delta_{k,\lambda_- l}\hskip2truecm k,l=0,1,\ldots,p-1
\end{equation}

The eigenstates of $U({\sf D}_{\sf A})$ are the multiplicative characters, $|\pi_0\rangle,|\pi_1\rangle,\ldots,|\pi_{p-1}\rangle$,
 of $\mathbb{F}_p^\ast=\{1,2,\ldots,p-1\}$, given by the expressions 
\begin{equation}
\label{multchar}
\begin{array}{l}
\displaystyle
\left\langle k|\pi_0\right\rangle=\delta_{k,0}\,\hskip1truecm k=0,1,2,\ldots,p-1\\
\end{array}
\end{equation}
\begin{equation}
\label{multchar1}
\begin{array}{l}
\displaystyle 
\left\langle 0|\pi_n\right\rangle = 0\\
\displaystyle 
\left\langle k|\pi_n\right\rangle = \frac{e^{\frac{2\pi\mathrm{i}n}{p-1}\mathrm{Ind}_g(k)}}{\sqrt{p-1}}
\end{array}
\end{equation}
where $k,n=1,2,\ldots,p-1$ and $\mathrm{Ind}_g(k)$ is the discrete logarithm of $k$ with respect to the base $g$, where $g$ is  a primitive element of $\mathbb{F}_p^\ast$; i.e. 

\begin{equation}
\label{primelem}
g^{\mathrm{Ind}_g(k)}\equiv k\,\mathrm{mod}\,p
\end{equation}
It follows that  $\mathrm{Ind}_g(k\cdot l)= \mathrm{Ind}_g(k) + \mathrm{Ind}_g(l)$. 

Having determined the eigenstates of $U({\sf D}_{\sf A})$, let us now provide the expressions of the eigenvalues. We remark that

\begin{equation}
\label{Uonpi}
\begin{array}{l}
\displaystyle
\left.U({\sf D}_{\sf A})|\pi_0\right\rangle = \left|\pi_0\right\rangle\\
\displaystyle
\left.U({\sf D}_{\sf A})|\pi_n\right\rangle = e^{-\frac{2\pi\mathrm{i}}{p-1}n\mathrm{Ind}_g(\lambda_+)}\left|\pi_n\right\rangle
\end{array}
\end{equation}
and can read off the eigenvalues of  $U({\sf A})$. The eigenvectors of $U({\sf A})$, $|\psi_n\rangle$, are given by 
\begin{equation}
\label{eigenevctorsUofA}
|\psi_n\rangle = U({\sf R})|\pi_n\rangle
\end{equation}

This calculation becomes effective using the explicit form of $U({\sf R})$, derived from the Weil representation (cf. appendix A.) 

The period of QACM is the period, $T(p)$, of the ACM and is, also, the order of the element(s), $\lambda_\pm$; since $\lambda_\pm$ are integers in $\mathbb{F}_p^\ast$, this order divides $p-1$, the order of $\mathbb{F}_p^\ast$. So there exists an integer, $\tau_p$, such that $p-1=\tau_p T(p)$. Therefore, $\mathrm{Ind}_g(\lambda_+)=\tau_p$. 

This is, precisely, the degeneracy of the eigenvalues of QACM, which are phases, $e^{\mathrm{i}\varepsilon_n}$. From the above we obtain

\begin{equation}
\label{eps_n}
\varepsilon_n = \frac{2\pi}{p-1}\tau_p n
\end{equation}
with $n=0,1,2,\ldots,p-1$. Since $n$ labels the eigenstates, too and $\tau_p/(p-1)=1/T(p)$ we can determine the degenerate eigenstates. 

With these tools we can write explicit expressions for the eigenstates, $|\psi_n\rangle$, 
\begin{equation}
\label{explicit_psi_n}
\langle k|\psi_n\rangle = \sum_{l=0}^{p-1}\langle k|U({\sf R})|l\rangle\langle l|\pi_n\rangle\Rightarrow\left\{
\begin{array}{l}
\displaystyle
\langle k|\psi_0\rangle = \langle k|U({\sf R})|0\rangle\\
\displaystyle
\langle k|\psi_n\rangle = \frac{1}{\sqrt{p-1}}\sum_{l=1}^{p-1}\langle k| U({\sf R})|l\rangle e^{\frac{2\pi\mathrm{i}}{p-1}n\tau_p}
\end{array}
\right.
\end{equation}
where $k=0,1,2,\ldots,p-1$, which will help to understand their chaotic properties. 

The degeneracies of the spectrum imply the existence of non--trivial conservation laws, that reduce the size of the attractor. As discussed in the last part of section 2,  we can determine explicitly, depending on p, the commutant of ${\sf A}$, $G({\sf A})$. 

This group is cyclic since we have chosen prime values for $p$, and, if its order is different from the period of ACM, there is a unique element, ${\sf B}$, which generates $G({\sf A})$. The corresponding quantum oeprator, $U({\sf B})$, generates the quantum conservation laws.

%================================================================================================= 
\section{ETH and the scrambling time bound for the QACM}\label{FibhorizonsSTB}
%==================================================================================================

Recently   there has been  a lot of activity around the question of the thermodynamics of closed  quantum systems~\cite{ETH}. 

An important role in this question has been assigned to the specific mechanisms of thermalization of various subsystems.

  A particularly interesting proposal is the Eigenstate Thermalization Hypothesis(ETH)~\cite{ETH}: 
  The time average of any observable of a  subsystem of a closed quantum system, which is, initially, in a pure state, for large times, converges to the thermal average of the observable, along with exponentially small corrections, $O(e^{-S})$, where $S$ is the entropy, defined by the thermal density matrix of the system. 
  
Since the total system is closed, it can be taken in a pure state and the temperature in the thermal density matrix is an effective temperature, defined by the energy average in the initial state of the total system.   

It has been shown~\cite{ETH} that a way to realize this hypothesis is to assume that the closed quantum system has a complete set of chaotic states in the specific sense that their probabilities  are sampled from a Gaussian pdf, while their phases are sampled from a flat pdf.

These are the basic premises for thermalization to be possible.  

In our particular chaotic, quantum, model of single particle scattering in the near horizon region of an extremal black hole, (some of) the basic ingredients of the ETH scenario  can be unambiguously identified. 

As we shall show below, the system of the black hole near horizon geometry, including the infalling wavepacket as a subsystem, has a complete set of chaotic eigenstates with, precisely, these properties. 

An important quantity that describes thermalization is the time required for thermalization. Recent studies have highlighted the relevance of the Hamiltonian dynamics of integrable and chaotic systems for determining the thermalization time. 

On the other hand, for  unitary, thermalization of the subsystem, it is interesting to study how--and if--the time required, is bounded, from below,  as a function of the entropy, for various physical systems~\cite{scrambling}. 

It has been conjectured that black hole horizons, considered, along with their probes, as closed quantum systems,  are the fastest scramblers~\cite{scrambling}. 

ETH, therefore, is  a  very interesting framework within which to discuss these issues. 

It has been recently conjectured that such a bound exists and that it is proportional to $(\beta/2\pi)\log S$, with $S$ the entropy and $\beta$ the (inverse) temperature and that black holes saturate it. 

In the AdS/CFT approach to this problem, we have the tools to study thermodynamics of gravitational backgrounds through the thermodynamics of the boundary, non--gravitational, conformal field theory. This is a consistent description of the thermodynamics of local, gravitational, observers, for which the observables are, indeed, defined, unambiguously, on the boundary. These are the sources for the boundary conformal field theory. 

Chaos is realized within the ETH, assuming that the dynamics of the closed quantum system is ergodic and mixing. This can be shown, using a random matrix description for the dynamics~\cite{ETH}.  For the case of the black hole, chaos can be described by shock wave geometries in the near horizon region.  For the thermodynamics of the black hole we expect to obtain a random ensemble of geometries and to determine its measure. Assuming ergodicity and a unitary, gravitational,  dynamics, this is equivalent to picking out a ``typical'' random geometry, as a gravitational background. 

Our approach to this end is  to use a particular arithmetic, namely modular,  discretization of the geometry, which,  while random, is  consistent with unitarity and holography.  

In the following we shall present arguments that support the statement that the QACM eigenfunctions do satisfy the assumptions of the ETH. 

The chaotic properties of the eigenstates can be traced back to the chaotic character of the discrete logarithm, $\mathrm{Ind}_g(l)$. The definition of chaos we shall adopt, which is the only one consistent with computational and algorithmic complexity, is that of  algorithmic chaos. 

The effective computation of the discrete logarithm is a classic example of a non--compressible algorithm, i.e. that cannot be done in polynomial time, with respect to the number of the input bits of $l$~\cite{arithmetic_geometry}. On the other hand, using quantum algorithms, Shor and others have shown that it can be reduced to polynomial complexity~\cite{shor}

 These considerations are consistent with the results of an old but very interesting paper of Ford {\em et al.} in ref.~\cite{quantum_maps} that shows  that the complexity of the QACM is $(\log\,N)^2$, in contrast with the classical one which is  $N$. This paper created a lot of discussion in the quantum chaos community (cf. the paper by Berry in Les Houches 1989~\cite{quantum_maps}).

The explicit expressions for the states $|\psi_n\rangle$ are sums of random phases, with fixed, complex, amplitudes. This leads, using the large number theorems, for $p\to\infty$, to Gaussian distributions of the state components. In the next section, we will provide numerical evidence for this claim and shall discuss some of the consequences regarding the randomness of the matrix QACM itself.

 As discussed in section 2 the mixing time for the classical ACM scales as the logarithm of the discretization parameter $N$ whenever $N$ takes values in the Fibonacci sequence.  The time required for unitary thermalization of a wavepacket (the scrambling time) is identified here with the mixing time $( t_\mathrm{scrambling}= t_\mathrm{mixing})$. 
This is so, because the period of the classical and the quantum ACMs coincide as a result of the construction of $U(A)$.
 
We find thus that the scrambling time of the QACM is proportional to $\log\,N$ when the dimension of the single particle Hilbert space, $N$,  takes values in the sequence of Fibonacci integers.
We recall  also  that the entropy $S$  of the  AdS$_2$  is proportional to $N$, the spatial extent of the geometry.
This leads to the saturation of the scrambling time bound, $t_\mathrm{mixing}=t_\mathrm{scrambling}\leq\log\,S$,   of Hayden-Preskill and Sekino-Susskind. 

The prefactor, which would be  the inverse of the temperature, here is to be replaced by the effective temperature of the closed quantum system, which depends on the complete set of its chaotic eigenstates,  according to the ETH scenario. For, while the Hawking temperature of the extremal black hole is zero, the chaotic dynamics of the extremal black hole microstates defines a consistently closed system, since the extremal black hole doesn't radiate. This point deserves a fuller analysis, that will be reported in future work. 

We  may  define finally the single particle scattering  S matrix as the evolution operator,  evaluated at half the period of the QACM.

 A consequence of the chaotic character of the QACM eigenstates is  that this matrix is random and completely delocalizes and scrambles initial Gaussian wavepackets.
 
\begin{figure}[thp]
\begin{center}
\includegraphics[scale=0.7]{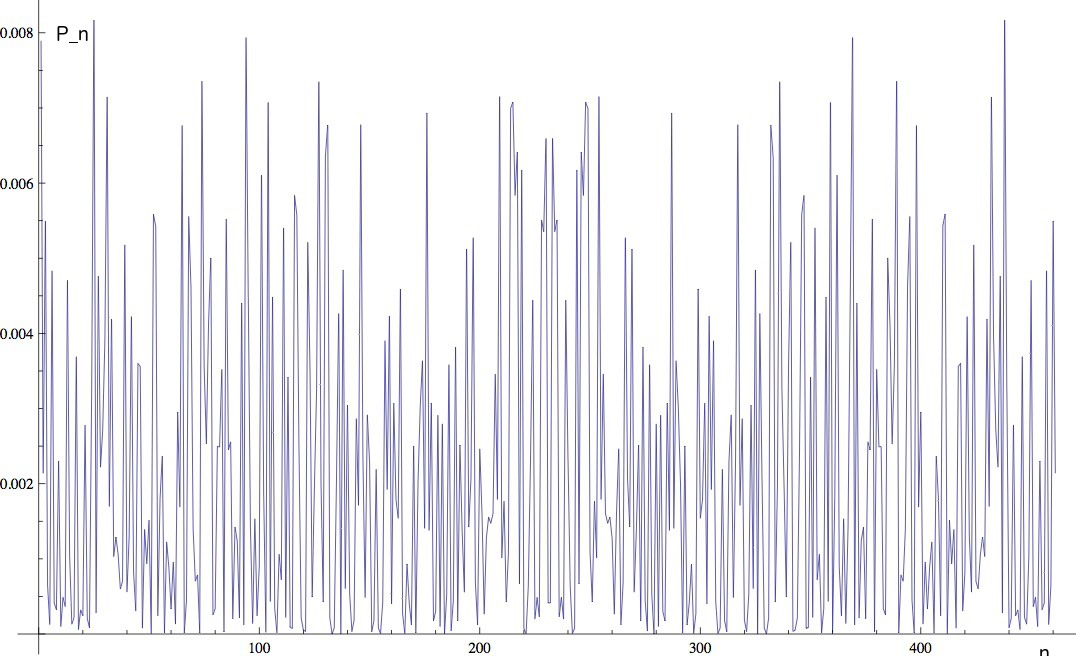}
\end{center}
\caption[]{The squared amplitude, $P_n\equiv |a_n|^2$, vs. the component label, $0\leq n\leq p-1$,  in the symmetric ground state (in the subspace of dimensionality $(p-1)/2$) of the quantum Arnol'd cat map, for $p=461$.}
\label{p=461GS}
\end{figure}

\begin{figure}[thp]
\begin{center}
\subfigure{\includegraphics[scale=0.7]{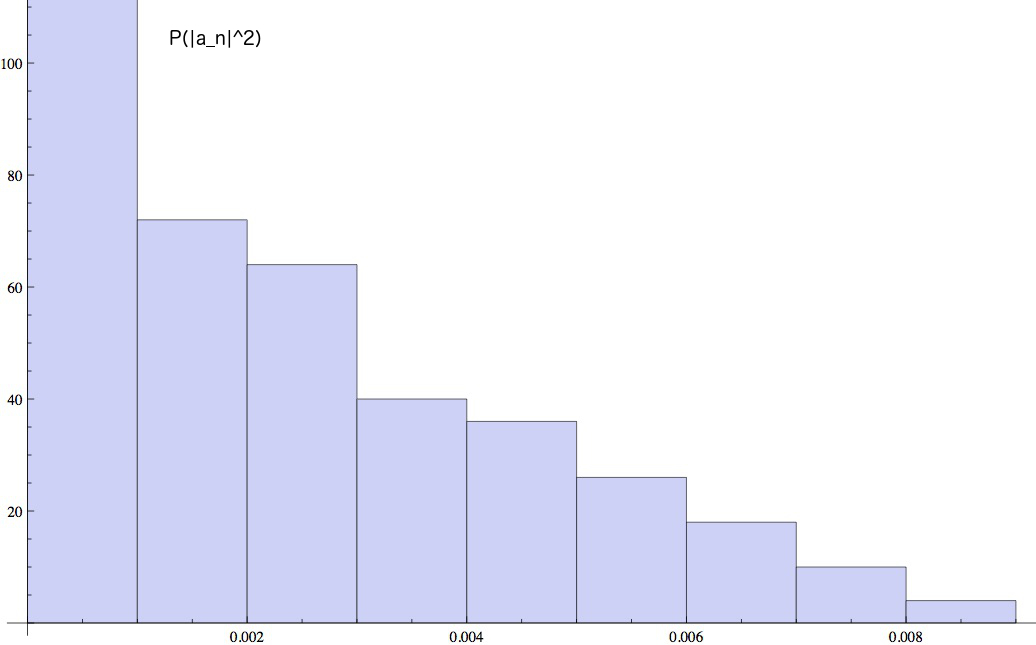}}
\subfigure{\includegraphics[scale=0.7]{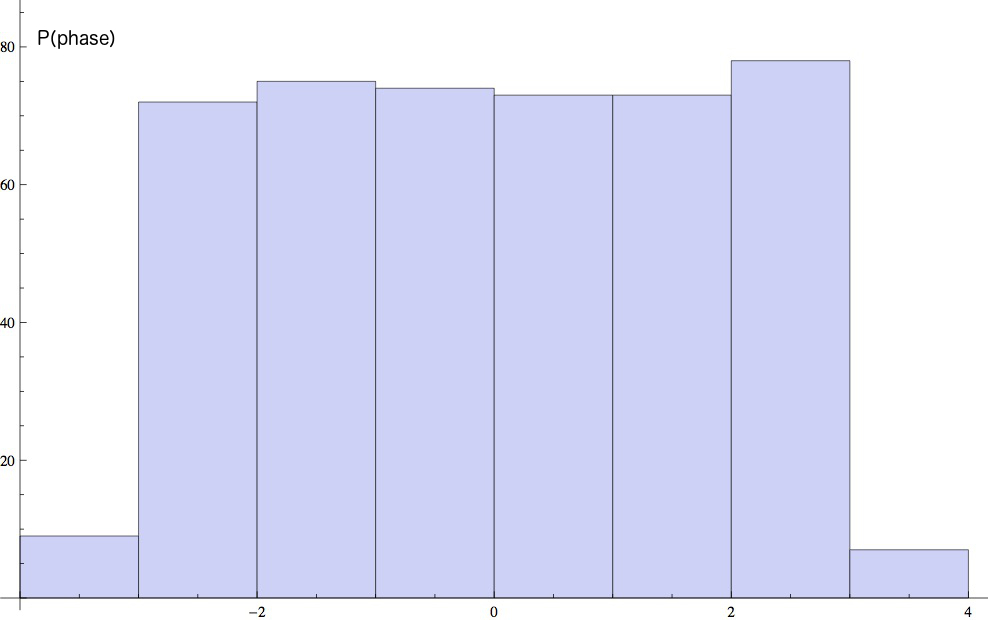}}
\end{center}
\caption[]{The binned  histograms  for the squared amplitude, $|a_n|^2$,  and for  the (relative) phase of the amplitude in the ground state, for $p=461$.}
\label{AmpPhases_p=461}
\end{figure}

Closing this section we shall present numerical support of our arguments for the chaotic nature of the eigenstates of QACM (cf. fig.~\ref{p=461GS} for the ground state for $p=461$). These results were obtained by using {\em Mathematica} codes. For $p=461$, 5 is a quadratic residue; but 461 isn't of the form $4k-1$; however it is possible to obtain the eigenvalues and eigenstates of the QACM numerically. In addition, for $p=461$, there exist two invariant subspaces, of dimensionalities $(p+1)/2=231$ and $(p-1)/2=230$. The, projective, representation, that is appropriate for AdS$_2$, as discussed above, is the second one. To highlight the symmetry of the ground state, we haven't projected onto the half length, but display the full length. 
From this figure it's possible to deduce that the probability distribution function (PDF) for the values of the amplitude squared is, indeed, Gaussian, and the phases are uniformly distributed cf. fig.~\ref{AmpPhases_p=461}. 

For Fibonacci dimensions of the Hilbert space of states, the period grows as $\log\,p$, which saturates the STB.  So, for large $p$, an exponentially small part  of spacetime   contributes to the chaotic dynamics, the rest are copies of it. This means that there is a large number of ``islands of chaos''.  ETH thus holds within each such island separately. 

This holds for the dynamics of the probes of the radial and temporal, AdS$_2$, geometry of the near horizon region; far from the horizon, at distances large compared to $\log\,p$, in units of the AdS$_2$ radius, the behavior becomes regular. The detailed crossover remains to be elucidated. 

Of course this numerical analysis is suggestive and will be completed in future work; but the big picture it defines is expected to be valid.

%================================================================================================
\section{Summary and conclusions }\label{Concl}
%===============================================================================================
In this work we proposed a toy model for the  chaotic scattering of single particle  wave packets in the modular discretization of  the radial AdS$_2$ space time  geometry of extremal (or nearly) extremal BHs.
In  recent  discussions  of the chaotic scattering the focus has been on the dynamics of the microscopic degrees of freedon on the streched horizon.  It is evident that,  although the dynamics of longitudinal and transverse scrambling will be different, the time  duration will be the same~\cite{silverstein}. 

We were able to discretize  the coset structure of  the geometry of this space time, by  introducing a modular invariant, infrared and ultraviolet cutoff.
We obtained an  AdS$_2$/CFT$_1$ holography and we provided the eigenstates and eigenvalues for the quantum chaotic Arnold cat map, as well as the single particle  S-matrix.

These eigenstates are chaotic in the sense of the eigenstate thermalization hypothesis. They create mixing and chaos for any infalling Gaussian wave packet.
 An interesting property of this model is that we can fix the dimension of the single particle Hilbert space of states, so that to saturate the scrambling time bound of Hayden-Preskill, Sekino and Susskind,  for an observer with time evolution defined by the QACM.
 
Our results   provide  a toy model mechanism to explain, how   the incoming information of a pure state can be scattered back as thermal radiation described by a density matrix (through the ETH scenario), while at the same time preserving unitarity in the single particle Hilbert space.

For future research along these lines, we think it would be interesting to  extend this toy model to the construction of the many particle or field theoretic chaotic scattering S-matrix, on the modular AdS$_2[N]$ geometry 
and  also investigate in  detail how the ETH works  in this case as well as to study the corresponding scrambling time. 

Finally describing the geometry with finite dimensional, $N=p^n$,  Hilbert  multi-qudit spaces, we provided  a framework of contact with the complexity theory of quantum algorithms and quantum circuits for the AdS/CFT correspondance~\cite{errorcorrectingAdSCFT}, since the finite unitary scattering  matrix can be written, as we shall show in a future work,  as a tensor product  of elementary qudit gates.

\vskip1.5truecm

{\bf Acknowledgements:} EGF and SN acknowledge the warm hospitality of the CERN Theory Division and in particular the organizers of the 2016 CERN Winter School for stimulating exchanges.  SN acknowledges the warm hospitality of the Institute of Nuclear and Particle Physics of the NCSR ``Demokritos''. 
 
 The research of EGF was partly implemented under the ``ARISTEIA-I'' action (Code no. 1612,  D.654) and title ``Holographic Hydrodynamics''  of the ``operational programme 
education and lifelong learning'' and is co-funded by the European Social Fund (ESF) and National Resources. 

The research of MA was supported in part by the grant MIS-448332-ORASY(NSRF 2007-13 ACTION, KRIPIS) of the European Regional Development Fund. 

\appendix

%====================================================================================
\section{ The Weil representation of PSL$_2[p]$ and the construction of the QACM eigenstates}\label{UofA}
%==================================================================================== 

Detailed references to this and the following appendix can be found in refs.~\cite{AFNAdS2CFT1} and~\cite{leoflo}.

The finite  Heisenberg--Weyl group  ${ HW}_p$, is defined as the set of $3\times 3$ matrices of the form
\begin{equation} 
 g(r,s,t)= \left(\begin{array}{ccc}1&0&0\\r&1&0\\t&s&1\end{array}\right)
\label{HeisGroup}
\end{equation}
where $r,s,t$ belong to $ \mathbb{Z}_p$ (integers modulo $p$), where the multiplication
of two elements is carried modulo $p$. 

\noindent 
When  $p$ is a prime integer   there is a unique $p$-dimensional  unitary irreducible and faithful representation of this group,
given  by the following matrices
 \begin{eqnarray} 
 \label{Jrs}
 J_{r,s,t}&=&\omega^t\,P^r Q^s
 \end{eqnarray}
where $\omega= e^{2\pi i/p}$, i.e. the $p^{th}$ primitive  root of unity and the matrices $P,Q$ are defined as
\begin{equation} 
\label{PandQ}
\begin{array}{lcl}
P_{kl}&=&\delta_{k-1,l}\\
Q_{kl}&=&\omega^{k}\delta_{kl}
\end{array}
\end{equation}
where $k,l=0,\dots, p-1$. 

It is to be  observed that, if  $\omega$  is replaced with  $\omega^k$,
for $k=1,2,...,p-1$ all the relations above remain intact.
Since $p$ is prime all the resulting representations are $p$-dimensional and inequivalent.

The matrices  $P,Q$  satisfy the fundamental Heisenberg commutation relation of
Quantum Mechanics in an exponentiated form
\begin{equation}
\label{WeylGroup} 
Q\,P =\omega\, P\,Q 
\end{equation}
In the above,  $Q$ represents the  position  operator on the circle  $\mathbb{Z}_p$  of the $p$ roots of unity
and $P$ the corresponding momentum operator. These two operators are related by the
diagonalising unitary matrix  $F$ of $P$,
\begin{equation}  
 \label{QPF}
QF =FP
\end{equation} 
so $F$ is the celebrated Discrete Fourier Transform  matrix 
\begin{equation}
\label{DFT}
  F_{kl}= \frac{1}{\sqrt{p}} \omega^{kl},\;{\rm  with }  \; k,l=0,\dots, p-1
  \end{equation}

An important subset of ${ HW}_p$ consists of  the magnetic translations 
\begin{equation} 
\label{JrsPQ}
J_{r,s}= \omega^{rs/2}P^rQ^s
\end{equation} 
with $r,s=0,\dots, p-1$.
These matrices are unitary ($J_{r,s}^{\dagger}=J_{-r,-s}$) and traceless, and  they form
a basis for the Lie algebra of $SL(p, \mathbb{C})$. They satisfy the important relation
\begin{equation} 
\label{JrsC}
J_{r,s}J_{r',s'} = \omega^{(r's-rs')/2}J_{r+r',s+s'}
\end{equation} 
This relation implies that the magnetic translations form a projective representation
of the translation group  $ \mathbb{Z}_p\times   \mathbb{Z}_p$. The factor of $1/2$ in the exponent
of (\ref{JrsC}) must be taken modulo $p$.

The $SL_2(p)$ appears here as the automorphism group of magnetic translations and this defines the Weil 
metaplectic representation. If we consider the action of an element 
\begin{equation}
\label{SL2pelement}
{\sf A}= \left(\begin{array}{cc}a&b\\c&d\end{array}\right)
\end{equation}
on the coordinates $(r,s)$ of the periodic torus  $ \mathbb{Z}_p\times   \mathbb{Z}_p$, this induces a unitary automorphism $U({\sf A})$ on the
magnetic translations, since the representation of Heisenberg group is unitary and irreducible,
\begin{equation}
\label{mtaplecticrep}
  U({\sf A})J_{r,s}U^{\dagger}({\sf A})= J_{r',s'} 
\end{equation}  
where $(r',s')$ are given by
\begin{eqnarray}
(r',s') &=& (r,s) \left(\begin{array}{cc}a&b\\c&d\end{array}\right)\label{JrsC1}
\end{eqnarray} 
This relation determines $U({\sf A})$ up to a phase and in the case of ${\sf A}\in \mathrm{SL}_2[p]$,  the phase can be
fixed to give  an exact (and not projective) unitary representation of SL$_2[p]$.

\noindent
The detailed formula of $U({\sf A})$ has been given by Balian and Itzykson~\cite{balian_itzykson}. Depending on the specific 
values of the $a,b,c,d$ parameters of the matrix ${\sf A}$, we distinguish the following cases:
\begin{equation} 
\label{BIcasesdelta}
\begin{array}{lcl}
\delta\ne 0: && U({\sf A})=\frac{\sigma(1)\sigma(\delta)}{p}\sum_{r,s}\omega^{\frac{br^2+(d-a)rs-cs^2}{2\delta}}J_{r,s}
\\
\delta = 0,\; b\ne 0:&& U({\sf A})=\frac{\sigma(-2b)}{\sqrt{p}}\sum_{s}\omega^{\frac{s^2}{2b}}J_{s(a-1)/b,s}
\\
\delta = b=0,\; c\ne 0:&& U({\sf A})=\frac{\sigma(2c)}{\sqrt{p}}\sum_{r}\omega^{-\frac{r^2}{2c}}P^r
\\
\delta = b=0= c= 0:&&U(1)=I
\end{array}
\end{equation}
where $\delta =2-a-d$ and $\sigma(a)$
is
the quadratic Gauss sum given by
\begin{equation} 
\label{QGauss}
\sigma(a) =\frac{1}{\sqrt{p}}\,\sum_{k=0}^{p-1}\omega^{ak^2} =(a|p)\times \left\{\begin{array}{cc}1&
{\rm for}\, p=4k+1\\ \mathrm{i}&{\rm for}\, p=4k-1\end{array}\right.
\end{equation}
while the Legendre symbol takes the values $(a|p)=\pm 1$ depending on whether $a$ is or is  not a square modulo $p$.

It is possible to perform explicitly the above Gaussian sums noticing that 
\begin{equation}
\label{Jrsrep}   
(J_{r,s})_{k,l} = \delta_{r,k-l} \omega^{\frac{k+l}2s}   
\end{equation}
where all indices take the values $k,l,r,s=0,\dots, p-1$. 
This has been done in ~\cite{athanasiu_floratos,fastqmaps}. In the case 
$\delta =2-a-d\ne 0\,\mathrm{mod}\,p$ and $c\ne 0\,\mathrm{mod}\,p$, the result is
\begin{equation}
\label{UAelms}
U({\sf A})_{k,l}=\frac{(-2c|p)}{\sqrt{p}}\,\times \left\{\begin{array}{c}1\\-\mathrm{i}\end{array}\right\}
\omega^{-\frac{ak^2-2kl+dl^2}{2c}}
\end{equation}

If $c\equiv 0\,\mathrm{mod}\,p$, then we transform the matrix ${\sf A}$ to one with $c\ne 0\,\mathrm{mod}\, p$. The  cases $\delta\equiv 0\,\mathrm{mod}\,p$ can be worked out easily using the expressions of the matrix elements of $J_{r,s}$, given in~(\ref{Jrs}).

It is interesting to notice that redefining $\omega$ to become $\omega^k$ for $k=1,2,...,p-1$, the matrix $U({\sf A})$ 
transforms to the matrix $U(A_k)$, where $A_k$  is the $2\times 2$ matrix
 ${\tiny A_k= \left(\begin{array}{cc}a&b k\\c/k&d\end{array}\right)}$, which 
belongs to the same conjugacy class with  $A$ as long as $k$ is  a quadratic residue.
If $k=p-1$  we pass from the representation  $U(A)$  to the complex conjugate one  $U(A)^*$.

The Weil representation presented above, provides the interesting result that the unitary matrix corresponding to the
$SL_2(p)$  element ${\tiny a=\left(\begin{array}{cc}0&-1\\1&0\end{array}\right)}$  is -up to a phase- the 
Discrete Finite Fourier Transform (\ref{DFT})
\[ U(a) =(-1)^{k+1}  i^n F    \]
where $n=0$ for $p=4k+1$ and $n=1$ for $p=4k-1$. 

The Fourier Transform matrix generates a fourth order abelian group with elements
\begin{equation} 
F,\; F^2=S, \;F^3=F^*,\; F^4=I\label{Fgroup}
\end{equation}

The matrix $S$ represents the element ${\tiny a^2=\left(\begin{array}{cc}-1&0\\0&-1\end{array}\right)}$ .
Its matrix elements are 
\begin{equation}
\label{UandS}
\begin{array}{lcl}
S_{k,l} &=&\delta_{k,-l},\;\; k,l=0,\dots p-1\\
U(a^2)_{k,l}&=&\mathrm{i}^{2n}S_{k,l} =(-)^n\delta_{k,-l},\;\; k,l=0,\dots p-1
\end{array}
\end{equation}
Because the action of $S$ on $J_{r,s}$ changes the signs of $r,s$, while  $\forall A\in SL_2(p)$ 
the unitary matrix $U(A)$  depends quadratically on $r,s$ in the sum (\ref{BIcasesdelta}), it turns out 
that $S$ commutes with all $U(A)$. Moreover, $S^2=I$ and we can construct two projectors 
\[ P_+=\frac 12( {I+S}),\; P_- = \frac 12 (I-S) \]
with dimensions of their invariant subspaces $\frac{p+1}2$ and $\frac{p-1}2$ correspondingly. So the Weil 
$p$-dimensional representation is the direct sum of two irreducible  
unitary representations 
\begin{equation}
\label{Uprojections}
U_+({\sf A})= U({\sf A}) P_+, \;  U_-({\sf A})=U(A)P_- 
\end{equation}

To obtain the block diagonal form of the above matrices $U_{\pm}(A)$, we rotate with the orthogonal 
matrix of the eigenvectors of $S$.
This $p$-dimensional orthogonal matrix, dubbed here $O_p$,
 can be obtained in a maximally symmetric form (along the diagonal as well as along the anti-diagonal) 
 using the eigenvectors of $S$ in the following order:
In the first $(p+1)/2$ columns we put the eigenvectors of $S$ of eigenvalue equal to 1,  and in the next
$(p-1)/2$ columns the eigenvectors of eigenvalue equal to $-1$ in the specific order given below:
\begin{equation}
\label{eigenvalues}
\begin{array}{lcl} 
({e_0})_k&=& \delta_{k0},\\
({e^+_j})_k&=&\frac{1}{\sqrt{2}}(\delta_{k,j}+\delta_{k,-j}),\; j=1,\dots, \frac{p-1}2\\
({e^-_j})_k&=&\frac{1}{\sqrt{2}}(\delta_{k,j}-\delta_{k,-j}),\; j=\frac{p+1}2,\dots, p
\end{array} 
\end{equation}
where $k=0,\dots, p-1$.

Different orderings of eigenvectors may lead to different  forms of the matrices $U_{\pm}(A)$.
The so obtained orthogonal matrix $O_p$ has the property 
\[  O_p^2 =I \]
due to its symmetric form.

The final block diagonal form of $U_{\pm}({\sf A})$ is obtained through an $O_p$ rotation
\begin{equation} 
\label{VM}
V_{\pm}({\sf A})= O_p U_\pm({\sf A}) O_p  
\end{equation}

In the following using the above material we shall provide the details of the construction of the eigenstates of the QACM. 

As we discussed in section~\ref{QACM}, the first step consists in diagonalizing the ACM in SL$_2[p]$ 
 and this can be done by an element ${\sf R}\in\mathrm{SL}_2[p]$ given by 
\begin{equation}
\label{Adiagonal}
{\sf R} = \left(\begin{array}{cc} \displaystyle \frac{a-1}{2} & \displaystyle -\frac{1}{2}\left(a^{-1}+1\right) \\
\displaystyle 1 & \displaystyle a^{-1}\end{array}\right)
\end{equation} 
where $a$ is the square root of 5 mod$\,p$. The matrix elements of $U({\sf A})$ and $U({\sf R})$ can be constructed explicitly, for any prime $p$, in particular for $p\equiv\,3\,\mathrm{mod}\,4$:
\begin{equation}
\label{UA}
\left[U({\sf A})\right]_{k,l}=\left\langle l|U({\sf A})|k\right\rangle=
\frac{-\mathrm{i}(-2|p)}{\sqrt{p}}\omega^{-\frac{1}{2}\left(k^2-2kl+2l^2\right)}
\end{equation}
and 
\begin{equation}
\label{UR}
\left[U({\sf R})\right]_{k,l}=\left\langle l|U({\sf R})|k\right\rangle=
\frac{-\mathrm{i}(-2|p)}{\sqrt{p}}\omega^{\frac{1}{4a}\left((5-a)k^2-4akl+2l^2\right)}
\end{equation}
where $k,l=0,1,2,\ldots,p-1$ and $\omega=\exp(2\pi\mathrm{i}/p)$. 

Plugging these expressions in  eq.~(\ref{explicit_psi_n}) we obtain the explicit forms of the QACM eigenstates, $\langle k|\psi_n\rangle$, $k,n=0,1,2,\ldots,p-1$:
\begin{equation}
\label{psi0eigen}
\left\langle k|\psi_{0}\right\rangle=\frac{-\mathrm{i}(-2|p)}{\sqrt{p}} \omega^{ - \frac{(5-a)k^{2}}{4a}}
\end{equation}

and 
\begin{equation}
\label{psineigen}
\left\langle k |\psi_{n}\right\rangle=
 \frac{-\mathrm{i}(-2|p)}{\sqrt{p(p-1)}} \sum_{\ell=1}^{p-1} \ 
e^{\frac{2\pi\mathrm{i}}{p-1}  n \ \mathrm{Ind}_{g} \ell}  \omega^{ - \frac{(5-a)k^{2} -4a
k  \ell + 2 \ell^{2} } { 4 a}} 
\end{equation}
We can project $U({\sf A})$ and its eigenstates onto the   two irreducible subspaces, of dimension $(p\pm 1)/2$, obtaining 
the block--diagonal forms $V_\pm({\sf A})$ 
\begin{equation}
\label{VpmA}
V_\pm({\sf A}) = O_p U({\sf A})\frac{1}{2}\left(I\pm S\right)O_p
\end{equation}
and   their corresponding eigenstates, $|\psi_n\rangle_\pm$ 
\begin{equation}
\label{psieigenpm}
|\psi_n\rangle_\pm = O_p\frac{1}{2}\left(I\pm S\right)|\psi_n\rangle 
\end{equation}
respectively. It can be checked that $V_\pm({\sf A})$ have the same period as $U({\sf A})$. 

These two irreducible representations of SL$2[p]$ are both appropriate for the torus, $\mathbb{T}^2$, not, however, for realizing the projective action of SL$_2[p]$ on AdS$_2$, as discussed in section~\ref{QACM}. We must choose that one of the two, which is, also, a (irreducible) representation of PSL$_2[p]$~\cite{rawnsley}.

We can obtain irreducible  representations of PSL$_2[p]$ from 
irreducible representations of SL$_2[p]$ in the following way:  The elements 
\begin{equation}
\label{PSL2p_a}
{\sf a}=\left(\begin{array}{cc}0&-1\\1&0\end{array}\right)
\end{equation} 
and 
\begin{equation}
\label{PSL2p_a2}
{\sf a}^2=\left(\begin{array}{cc}-1&0\\0&-1\end{array}\right)
\end{equation} 
of SL$_2[p]$
have representatives
\begin{equation}
\label{Ua}
U({\sf a}) = (-1)^{k+1}  \mathrm{i}^n F 
\end{equation}
and 
\begin{equation}
\label{Ua2}
U({\sf a}^2)= (-1)^n S
\end{equation}

where $n=0$ for $p=4k+1$ and $n=1$ for $p=4k-1$. 

For PSL$_2[p]$ the element ${\sf a}^2=-I$ is identified with the identity matrix, $I$. Therefore, we should choose, among the two irreducible representations of SL$_2[p]$, of dimension $(p+1)/2$ and $(p-1)/2$, that one, for which $U({\sf a}^2)=I$.  

We can easily check  that this happens  for the 
$\frac{p+1}{2}$ dimensional representation,  when $p\equiv\,1\,\mathrm{mod}\,4$, and for the $\frac{p-1}{2}-$dimensional one, when $p\equiv\,3\,\mathrm{mod}\,4$. In our construction we found it simpler to work with primes  of the latter form, therefore the representation is that defined by $V_-({\sf A})$, with eigenstates, $|\psi_n\rangle_-$. 

The corresponding eigenvalues, $\varepsilon_n$, in eq.~(\ref{eps_n}), have index ranging in $n=(p+1)/2,\ldots,p-1$.


\begin{thebibliography}{99}
\bibitem{hooft}
G.~'t Hooft,
"Diagonalizing the black hole Information retrieval process"
[arXiv: 1509.01695[gr-qc]];
``The Quantum Black Hole as a Hydrogen Atom: Microstates Without Strings Attached,''
  [arXiv:1605.05119 [gr-qc]];
  %%CITATION = ARXIV:1605.05119;%%
 " Black hole unitarity and antipodal entanglement"
[arXiv: 1601.03447 [gr-qc]].
%%CITATION = ARXIV:1601.03447;%%

\bibitem{strom}
A.~Strominger and A.~Zhiboedov,
  ``Gravitational Memory, BMS Supertranslations and Soft Theorems,''
  JHEP {\bf 1601} (2016) 086
  doi:10.1007/JHEP01(2016)086
  [arXiv:1411.5745 [hep-th]].
  %%CITATION = doi:10.1007/JHEP01(2016)086;%%
  
  S.~W.~Hawking, M.~J.~Perry and A.~Strominger,
  ``Soft Hair on Black Holes,''
  Phys.\ Rev.\ Lett.\  {\bf 116} (2016) no.23,  231301
  doi:10.1103/PhysRevLett.116.231301
  [arXiv:1601.00921 [hep-th]].
  %%CITATION = doi:10.1103/PhysRevLett.116.231301;%%

J.~Ellis, N.~E.~Mavromatos and D.~V.~Nanopoulos,
  ``$W_\infty$ Algebras, Hawking Radiation and Information Retention by Stringy Black Holes,''
  Phys.\ Rev.\ D {\bf 94} (2016) no.2,  025007
  doi:10.1103/PhysRevD.94.025007
  [arXiv:1605.01653 [hep-th]].
  %%CITATION = doi:10.1103/PhysRevD.94.025007;%%
\bibitem{hardsoft}
A.~Strominger,
  ``Black Hole Information Revisited,''
  arXiv:1706.07143 [hep-th].
  %%CITATION = ARXIV:1706.07143;%%
  
  D.~Carney, L.~Chaurette, D.~Neuenfeld and G.~W.~Semenoff,
  ``Infrared quantum information,''
  arXiv:1706.03782 [hep-th].
  %%CITATION = ARXIV:1706.03782;%% 
  
\bibitem{barrabes_etal} C. Barrab\`es, V. Frolov and R. Parentani, ``Stochastically fluctuating black hole geometry, Hawking radiation and the trans--Planckian problem'', {\sl Phys. Rev.} {\bf D62} (2000) 044020.
doi:10.1103/PhysRevD.62.044020
  [gr-qc/0001102].
  %%CITATION = doi:10.1103/PhysRevD.62.044020;%%

\bibitem{papadodimas} 
S.~Banerjee, J.~W.~Bryan, K.~Papadodimas and S.~Raju,
  ``A toy model of black hole complementarity,''
  JHEP {\bf 1605} (2016) 004
  doi:10.1007/JHEP05(2016)004
  [arXiv:1603.02812 [hep-th]].
  %%CITATION = doi:10.1007/JHEP05(2016)004;%%
  
K.~Papadodimas and S.~Raju,
  ``An Infalling Observer in AdS/CFT,''
  JHEP {\bf 1310} (2013) 212
  %doi:10.1007/JHEP10(2013)212
  [arXiv:1211.6767 [hep-th]].  
%%CITATION = doi:10.1007/JHEP10(2013)212;%%    
  
\bibitem{stanford_susskind}  
D.~Stanford and L.~Susskind,
  ``Complexity and Shock Wave Geometries,''
  Phys.\ Rev.\ D {\bf 90} (2014) no.12,  126007
  doi:10.1103/PhysRevD.90.126007
  [arXiv:1406.2678 [hep-th]].
  %%CITATION = doi:10.1103/PhysRevD.90.126007;%%

\bibitem{shenker_stanford}
S.~H.~Shenker and D.~Stanford,
  ``Multiple Shocks,''
  JHEP {\bf 1412} (2014) 046
  doi:10.1007/JHEP12(2014)046
  [arXiv:1312.3296 [hep-th]].
  %%CITATION = doi:10.1007/JHEP12(2014)046;%%  
  
S.~H.~Shenker and D.~Stanford,
  ``Black holes and the butterfly effect,''
  JHEP {\bf 1403} (2014) 067
  doi:10.1007/JHEP03(2014)067
  [arXiv:1306.0622 [hep-th]].
  %%CITATION = doi:10.1007/JHEP03(2014)067;%%
 
 J.~Maldacena, S.~H.~Shenker and D.~Stanford,
  ``A bound on chaos,''
  JHEP {\bf 1608} (2016) 106
 % doi:10.1007/JHEP08(2016)106
  [arXiv:1503.01409 [hep-th]].  
  %%CITATION = doi:10.1007/JHEP08(2016)106;%%

   A.~L.~Fitzpatrick and J.~Kaplan,
  ``A Quantum Correction To Chaos,''
  JHEP {\bf 1605} (2016) 070
%  doi:10.1007/JHEP05(2016)070
  [arXiv:1601.06164 [hep-th]].  
%%CITATION = doi:10.1007/JHEP05(2016)070;%%
  

  P.~Caputa, J.~Simón, A.~\v{S}tikonas, T.~Takayanagi and K.~Watanabe,
  ``Scrambling time from local perturbations of the eternal BTZ black hole,''
  JHEP {\bf 1508} (2015) 011
  doi:10.1007/JHEP08(2015)011
  [arXiv:1503.08161 [hep-th]].
  %%CITATION = doi:10.1007/JHEP08(2015)011;%%    
  
  P.~Padmanabhan, S.~J.~Rey, D.~Teixeira and D.~Trancanelli,
  ``Supersymmetric many-body systems from partial symmetries — integrability, localization and scrambling,''
  JHEP {\bf 1705} (2017) 136
  doi:10.1007/JHEP05(2017)136
  [arXiv:1702.02091 [hep-th]].
  %%CITATION = doi:10.1007/JHEP05(2017)136;%%
  
\bibitem{mezei_stanford}
M.~Mezei and D.~Stanford,
  ``On entanglement spreading in chaotic systems,''
  arXiv:1608.05101 [hep-th].
  %%CITATION = ARXIV:1608.05101;%%  
  
\bibitem{polchinski}
J.~Polchinski,
  ``Chaos in the black hole S-matrix,''
  arXiv:1505.08108 [hep-th].
  %%CITATION = ARXIV:1505.08108;%% 
  
   J.~Polchinski,
  ``The Black Hole Information Problem,''
  arXiv:1609.04036 [hep-th].
  %%CITATION = ARXIV:1609.04036;%% 

\bibitem{sussk}
A.~R.~Brown, L.~Susskind and Y.~Zhao,
  ``Quantum Complexity and Negative Curvature,''
  arXiv:1608.02612 [hep-th].
  %%CITATION = ARXIV:1608.02612;%%  
  
J.~Engelsöy, T.~G.~Mertens and H.~Verlinde,
  ``An investigation of AdS$_{2}$ backreaction and holography,''
  JHEP {\bf 1607} (2016) 139
%  doi:10.1007/JHEP07(2016)139
  [arXiv:1606.03438 [hep-th]].
  

\bibitem{BFKS} 
T.~Banks, W.~Fischler, I.~R.~Klebanov and L.~Susskind,
  ``Schwarzschild black holes from matrix theory,''
  Phys.\ Rev.\ Lett.\  {\bf 80} (1998) 226
  doi:10.1103/PhysRevLett.80.226
  [hep-th/9709091].
  %%CITATION = doi:10.1103/PhysRevLett.80.226;%% 
  
\bibitem{polchinski08}
N.~Iizuka, T.~Okuda and J.~Polchinski,
  ``Matrix Models for the Black Hole Information Paradox,''
  JHEP {\bf 1002} (2010) 073
  doi:10.1007/JHEP02(2010)073
  [arXiv:0808.0530 [hep-th]].
  %%CITATION = doi:10.1007/JHEP02(2010)073;%%   
  
\bibitem{avery}
S.~G.~Avery,
  ``Qubit Models of Black Hole Evaporation,''
  JHEP {\bf 1301} (2013) 176
  doi:10.1007/JHEP01(2013)176
  [arXiv:1109.2911 [hep-th]].
  %%CITATION = doi:10.1007/JHEP01(2013)176;%%  


\bibitem{magan} 
J.~M.~Magan,
  ``Black holes as random particles: entanglement dynamics in infinite range and matrix models,''
  arXiv:1601.04663 [hep-th].
  %%CITATION = ARXIV:1601.04663;%%
  

  
\bibitem{AFNAdS2CFT1} M.~Axenides, E.~G.~Floratos and S.~Nicolis,
  ``Modular discretization of the AdS$_{2}$/CFT$_{1}$ holography,''
  JHEP {\bf 1402} (2014) 109
  [arXiv:1306.5670 [hep-th]].
  %%CITATION = ARXIV:1306.5670;%%

 M.~Axenides, E.~Floratos and S.~Nicolis,
  ``Chaotic Information Processing by Extremal Black Holes,''
  Int.\ J.\ Mod.\ Phys.\ D {\bf 24} (2015) no.09,  1542012
  doi:10.1142/S0218271815420122
  [arXiv:1504.00483 [hep-th]].
  %%CITATION = doi:10.1142/S0218271815420122;%%
   
 E. Floratos, “The chaotic eigenstate hypothesis and fast scrambling on BH horizons:
A quantum Arnol’d cat map toy model”, 8th Crete regional meeting on string theory,
Nafplion 2015, http://hep.physics.uoc.gr/mideast8/talks/tuesday/Floratos.pdf

E. Floratos, ``Quantum complexity and chaotic dynamics on extremal black hole horizons'',
9th Crete regional meeting on string theory,
http://hep.physics.uoc.gr/mideast9/talks/tuesday/floratos.pdf
   
\bibitem{townsendetal} 
P.~Claus, M.~Derix, R.~Kallosh, J.~Kumar, P.~K.~Townsend and A.~Van Proeyen,
  ``Black holes and superconformal mechanics,''
  {\sl Phys.\ Rev.\ Lett.}  {\bf 81} (1998) 4553
  [arXiv:hep-th/9804177].
  %%CITATION = HEP-TH/9804177;%%

\bibitem{arnoldcatmapbooks} 
V. I. Arnol'd and A. Avez, {\em Ergodic problems in classical mechanics}, Benjamin, N. Y. (1968).

G. M. Zaslavsky, {\em Hamiltonian chaos and fractional dynamics}, Oxford University Press (2008). 

Ya. G. Sinai, {\em Topics in ergodic theory}, Princeton University Press (1994). 


\bibitem{quantum_maps} 

M. V. Berry, N. L. Balazs, M. Tabor and A Voros, ``Quantum maps'', {\sl Annals of Physics} {\bf 122} (1979) 26.  

J. H. Hannay and M. V. Berry, ``Quantization of linear maps on a torus'', {\sl Physica} {\bf D1} (1980) 267.

J. Ford, G. Mantica and G. H. Ristow, ``The Arnol'd cat: failure of the correspondence principle'', {\sl Physica} {\bf D50} (1991) 493.

M. V. Berry, ``Some quantum to classical asymptotics'', {\em Les Houches Summer School 1989}, Elsevier (1991). 

\bibitem{knabe} S. Knabe, ``On the quantization of Arnol'd's cat'', {\sl J. Phys. A: Math. Gen.} {\bf 23} (1990) 2013. 

\bibitem{floratos89} 
E.~G.~Floratos,
  ``The Heisenberg-Weyl Group On The $\mathbb{Z}_n\times  \mathbb{Z}_n$ Discretized Torus Membrane,''
  {\sl Phys.\ Lett.}  {\bf B228} (1989) 335.
%%CITATION = doi:10.1016/0370-2693(89)91555-4;%%


    
\bibitem{athanasiu_floratos} 
G. G. Athanasiu and E. G. Floratos,
``Coherent states in finite quantum mechanics,''
  {\sl Nucl.\ Phys.}  {\bf B425} (1994) 343.
  
\bibitem{fastqmaps}   
G.~G.~Athanasiu, E.~G.~Floratos and S.~Nicolis,
  ``Fast quantum maps,''
  {\sl J.\ Phys.}   {\bf A31} (1998) L655
  [arXiv:math-ph/9805012]
  %%CITATION = doi:10.1088/0305-4470/31/38/001;%%

\bibitem{arithmetic_geometry} 
E. Artin, ``Alg\`ebre g\'eom\'etrique'', Eds. J. Gabay (1996). 

D. Bressoud and S. Wagon, ``Computational number theory'', Key college publishing (2000). 

N. Koblitz, ``A course in number theory and cryptography'', Springer-Verlag, New York(1994). 

V. I. Arnol'd, ``Dynamics, statistics and projective geometry of Galois fields'', Cambridge university press (2011). 

\bibitem{ETH} M. V. Berry,  ``Regular and irregular semiclassical wave functions'', J. Phys. A: Math. Gen.  10, 2083-91 (1977).

J. M. Deutsch, ``Quantum statistical mechanics in a closed system'', Phys. Rev.  A43 (1991) 2046.

M.  Srednicki, ``Quantum Chaos and Statistical Mechanics'', 	10.1111/j.1749-6632.1995.tb39017.x, Ann. N. Y. Acad. Sc. 755, (1995) 757.  	
[arXiv:cond-mat/9406056] %%CITATION = arXiv:cond-mat/9406056;%%

J. M. Magan, ``Random free fermions: An analytical example of eigenstate thermalization'', Phys. Rev. Lett. 116, 030401 (2016). 
arXiv:1508.05339v2 [quant-ph] %%CITATION = doi:10.1103/PhysRevLett.116.030401;%%

 B.~Craps, O.~Evnin and K.~Nguyen,
  ``Matrix Thermalization,''
  JHEP {\bf 1702} (2017) 041
%  doi:10.1007/JHEP02(2017)041
  [arXiv:1610.05333 [hep-th]].   

R.~H\"ubener, Y.~Sekino and J.~Eisert,
  ``Equilibration in low-dimensional quantum matrix models,''
  JHEP {\bf 1504} (2015) 166
  doi:10.1007/JHEP04(2015)166
  [arXiv:1403.1392 [quant-ph]].  
  
J.~Sonner and M.~Vielma,
  ``Eigenstate thermalization in the Sachdev-Ye-Kitaev model,''
  arXiv:1707.08013 [hep-th].
  %%CITATION = ARXIV:1707.08013;%%  
  
D.~C.~Brody, D.~W.~Hook and L.~P.~Hughston,
  ``Unitarity, ergodicity, and quantum thermodynamics,''
  J.\ Phys.\ A {\bf 40} (2007) F503
  doi:10.1088/1751-8113/40/26/F01
  [quant-ph/0702009 [QUANT-PH]].
  %%CITATION = doi:10.1088/1751-8113/40/26/F01;%%

\bibitem{falk_dyson} F. J. Dyson and H. Falk, ``Period of a discrete cat mapping'', {\sl Am. Math. Monthly} {\bf 99} (1992) 603.

 
\bibitem{quantum_ergodicity}
M D Esposti and S Isola, ``Distribution of closed orbits for linear automorphisms of tori'', {\sl Nonlinearity} {\bf 8} (1995) 821.

P. Kurlberg and Z. Rudnick, ``On quantum ergodicity for linear maps of the torus'',  Comm. Math. Phys. {\bf 222} (2001) 201. 
 [arXiv:math/9910145v1 [math.NT]] %%CITATION = arXiv:math/9910145v1;%%
 
 P. Kurlberg and Z. Rudnick, ``On the distribution of matrix elements for the quantum cat map'', Ann. Math. {\bf 161} (2005) 489.
 [arXiv:math/0302277v3 [math.NT]] %%CITATION = arXiv:math/0302277v3;%%
 
 
F. Faure, S. Nonnenmacher, S. De Bievre, ``Scarred eigenstates for quantum cat maps of minimal periods'', 
Commun.Math.Phys. 239, 449-492 (2003).  [arXiv:nlin/0207060v2 [nlin.CD]] %%CITATION = arXiv:nlin/0207060v2;%%

S. Zelditch, ``Recent developments in mathematical Quantum Chaos'',  Current Developments in Mathematics 2009, p. 115- 202. 
[arXiv:0911.4312v1 [math.AP]] %%CITATION = arXiv:0911.4312v1;%%


\bibitem{scrambling} 
P.~Hayden and J.~Preskill,
  ``Black holes as mirrors: Quantum information in random subsystems,''
  JHEP {\bf 0709} (2007) 120
  doi:10.1088/1126-6708/2007/09/120
  [arXiv:0708.4025 [hep-th]].
  %%CITATION = doi:10.1088/1126-6708/2007/09/120;%%

Y.~Sekino and L.~Susskind,
  ``Fast Scramblers,''
  JHEP {\bf 0810} (2008) 065
  doi:10.1088/1126-6708/2008/10/065
  [arXiv:0808.2096 [hep-th]].
  %%CITATION = doi:10.1088/1126-6708/2008/10/065;%%  
  
J. L. F. Barbon, J. M. Magan, ``Fast Scramblers And Ultrametric Black Hole Horizons'', J. High Energ. Phys. (2013) 2013: 163. [ arXiv:1306.3873v2 [hep-th]] %%CITATION = doi:10.1007/JHEP11(2013)163;%%

S.~Pramodh and V.~Sahakian,
  ``From Black Hole to Qubits: Evidence of Fast Scrambling in BMN theory,''
  JHEP {\bf 1507} (2015) 067
%  doi:10.1007/JHEP07(2015)067
  [arXiv:1412.2396 [hep-th]].  

N.~Lashkari, D.~Stanford, M.~Hastings, T.~Osborne and P.~Hayden,
  ``Towards the Fast Scrambling Conjecture,''
  JHEP {\bf 1304} (2013) 022
  doi:10.1007/JHEP04(2013)022
  [arXiv:1111.6580 [hep-th]].  
  %%CITATION = doi:10.1007/JHEP04(2013)022;%%

J.~L.~F.~Barbon and J.~M.~Magan,
  ``Fast Scramblers Of Small Size,''
  JHEP {\bf 1110} (2011) 035
  %doi:10.1007/JHEP10(2011)035
  [arXiv:1106.4786 [hep-th]].  
  

\bibitem{witten_susskind} L.~Susskind and E.~Witten,
  ``The Holographic bound in anti-de Sitter space,''
  hep-th/9805114.
  %%CITATION = HEP-TH/9805114;%%

\bibitem{bh_entropy}
A.~Sen,
  ``State Operator Correspondence and Entanglement in $AdS_2/CFT_1$,''
  Entropy {\bf 13} (2011) 1305
  doi:10.3390/e13071305
  [arXiv:1101.4254 [hep-th]].
  %%CITATION = doi:10.3390/e13071305;%%

A.~Dabholkar and S.~Nampuri,
  ``Quantum black holes,''
  Lect.\ Notes Phys.\  {\bf 851} (2012) 165
  doi:10.1007/978-3-642-25947-0$_5$
  [arXiv:1208.4814 [hep-th]].
  %%CITATION = doi:10.1007/978-3-642-25947-0_5;%%
  
T.~Azeyanagi, T.~Nishioka and T.~Takayanagi,
  ``Near Extremal Black Hole Entropy as Entanglement Entropy via AdS(2)/CFT(1),''
  Phys.\ Rev.\ D {\bf 77} (2008) 064005
  doi:10.1103/PhysRevD.77.064005
  [arXiv:0710.2956 [hep-th]].
  %%CITATION = doi:10.1103/PhysRevD.77.064005;%%
    
\bibitem{shor} P. Shor, ``Polynomial-Time Algorithms for Prime Factorization and Discrete Logarithms on a Quantum Computer'', SIAM J.Sci.Statist.Comput. 26 (1997) 1484 [arXiv:quant-ph/9508027] 
 %%CITATION = doi:10.1137/S0097539795293172;%%

\bibitem{silverstein}
M.~Dodelson and E.~Silverstein,
  ``Longitudinal nonlocality in the string S-matrix,''
  arXiv:1504.05537 [hep-th].
  %%CITATION = ARXIV:1504.05537;%%
  
 L.~Susskind and J.~Lindesay,
  ``An introduction to black holes, information and the string theory revolution: The holographic universe,''
  Hackensack, USA: World Scientific (2005) 183 p  
  
\bibitem{errorcorrectingAdSCFT} 
 F.~Pastawski, B.~Yoshida, D.~Harlow and J.~Preskill,
  ``Holographic quantum error-correcting codes: Toy models for the bulk/boundary correspondence,''
  JHEP {\bf 1506} (2015) 149
  doi:10.1007/JHEP06(2015)149
  [arXiv:1503.06237 [hep-th]].
  %%CITATION = doi:10.1007/JHEP06(2015)149;%%
  

\bibitem{rawnsley} 
John Rawnsley, ``On the universal covering group of the real symplectic group'', 
Journal of Geometry and Physics, Volume 62, Issue 10, October 2012, Pages 2044-2058, ISSN 0393-0440, http://dx.doi.org/10.1016/j.geomphys.2012.05.009.  

\bibitem{leoflo}  E.~G.~Floratos and G.~K.~Leontaris,
  ``Discrete Flavour Symmetries from the Heisenberg Group,''
  Phys.\ Lett.\ B {\bf 755} (2016) 155
  doi:10.1016/j.physletb.2016.02.007
  [arXiv:1511.01875 [hep-th]].
  %%CITATION = doi:10.1016/j.physletb.2016.02.007;%%

\bibitem{balian_itzykson} R. Balian and C. Itzykson, ``Observations on finite quantum mechanics'', {\em C. R. Acad. Sc. Paris} {\bf 303} I (1986) 773.



\end{thebibliography}
\end{document}